\documentclass[runningheads]{llncs}
\usepackage[T1]{fontenc}
\usepackage{graphicx, subcaption, url}
\usepackage{caption, enumitem}
\usepackage{amsmath, amssymb}
\begin{document}
\title{Maximum Entropy Models for Unimodal Time Series: Case Studies of Universe 25 and St. Matthew Island}
\titlerunning{Maximum Entropy Models for Unimodal Time Series}
%
\author{Sabin Roman\inst{1}}
%
%
\institute{Department of Knowledge Technologies, Jo\v{z}ef Stefan Institute\\
\email{sabin.roman@ijs.si}\\
ORCID: \url{https://orcid.org/0000-0002-7513-3479}}
\maketitle              
\begin{abstract}

We present a maximum entropy modeling framework for unimodal time series: signals that begin at a reference level, rise to a single peak, and return. Such patterns are commonly observed in ecological collapse, population dynamics, and resource depletion. Traditional dynamical models are often inapplicable in these settings due to limited or sparse data, frequently consisting of only a single historical trajectory. In addition, standard fitting approaches can introduce structural bias, particularly near the mode, where most interpretive focus lies. Using the maximum entropy principle, we derive a least-biased functional form constrained only by minimal prior knowledge, such as the starting point and estimated end. This leads to analytically tractable and interpretable models. We apply this method to the collapse of the Universe 25 mouse population and the reindeer crash on St. Matthew Island. These case studies demonstrate the robustness and flexibility of the approach in fitting diverse unimodal time series with minimal assumptions. We also conduct a cross-comparison against established models, including the Richards, Skewnormal, and Generalized Gamma functions. While models typically fit their own generated data best, the maximum entropy models consistently achieve the lowest off-diagonal root-mean-square losses, indicating superior generalization. These results suggest that maximum entropy methods provide a unifying and efficient alternative to mechanistic models when data is limited and generalization is essential.

\end{abstract}
\section{Introduction}

Unimodal patterns, characterized by a single dominant peak, frequently arise in diverse natural, social, and economic systems, often signaling phases of growth followed by decline or reflecting intrinsic distributional structures. Understanding and modeling these phenomena is essential for predicting critical transitions and characterizing complex systems. While two, three or higher dimensional time series can be effectively modeled through dynamical systems such as differential equations or chaotic maps \cite{roman2021historical,roman2023theories,roman2022master,roman2023global,roman2019growth}, or agent-based models \cite{bertolotti2024balancing}, in the case of sparse data for a single variable time series these methods can prove overly complex, under-constrained, and prone to overfitting - making simple, shape-constrained models like the maximum entropy functions presented here a more robust and interpretable alternative.

In this work, we develop and apply maximum entropy models to characterize unimodal data, encompassing both dynamic time series and static distributions. We illustrate the approach through two prominent case studies: the population dynamics of Universe 25 and the reindeer population collapse on St. Matthew Island. Our results demonstrate that maximum entropy formulations effectively capture the essential features of unimodal rise-and-fall behavior as well as stable unimodal distributions, providing a unified statistical framework. This approach highlights the potential of entropy-based models to serve as universal descriptors across ecological, resource, and socio-economic systems.

\section{Methods}

The intersection of machine learning and time series modeling has witnessed significant advances through the development of structured and multi-target learning approaches. Techniques such as predictive clustering trees (PCTs) and model trees have emerged as powerful tools for capturing complex temporal dynamics, especially in settings involving multivariate or structured outputs \cite{osojnik2025isoup,stevanoski2024change,stoimchev2024semi}. These methods excel in modeling tasks where temporal data is not only univariate but involves interconnected outputs evolving over time - scenarios often encountered in environmental and ecological domains \cite{ispirova2024msgen,andonovikj2024survival,levatic2024semi}.

Multi-target regression has been particularly influential for time series applications that require simultaneous prediction of several dependent variables \cite{kostovska2022explainable,stoimchev2023deep,petkovic2023clusplus}. In environmental monitoring, for instance, forecasting the simultaneous progression of temperature, humidity, and pollutant concentrations benefits from algorithms capable of exploiting inter-target correlations. Model trees tailored for multi-target prediction allow for accurate and interpretable models, especially when combined with hierarchical or taxonomically structured outputs, as seen in ecological and biological time series \cite{petkovic2022relational,bogatinovski2022comprehensive,novak2022occurrence}.

Symbolic regression, particularly via evolutionary approaches, has further extended the ability to discover functional forms underlying temporal trends. These methods are capable of capturing the underlying generative processes of time series data, producing models that are both predictive and interpretable \cite{omejc2024probabilistic,brence2023dimensionally,gec2022discovery,radinja2021automated,tanevski2020combinatorial,simidjievski2020equation,lukvsivc2019meta,tolovski2019towards,tanevski2017process,simidjievski2016modeling,KompareDzeroski1994}. Such capabilities are valuable in scientific discovery settings where understanding the temporal dynamics is as critical as accurate forecasting.

In real-world applications, hybrid modeling frameworks that integrate empirical data with domain knowledge have shown substantial promise. These approaches are particularly suited for time series exhibiting regular patterns, such as seasonal unimodal curves in phenology, hydrology, and agriculture. The use of hybrid models supports generalization across systems with similar structural characteristics but differing quantitative parameters.

Furthermore, the emphasis on interpretable modeling has driven the adoption of rule-based and tree-based approaches for temporal analysis. These models not only achieve high accuracy but also support model transparency, enabling domain experts to validate and refine predictive insights, which is essential in high-stakes applications like environmental risk assessment or medical prognosis.

While structured and multi-target machine learning approaches have greatly enhanced our ability to model complex temporal systems, they often involve considerable algorithmic complexity and require large volumes of data. Tree-based models and symbolic regression frameworks, for instance, offer flexibility and interpretability, especially in multivariate or relational contexts. However, their application to relatively simple univariate time series—particularly those characterized by a single peak—can introduce unnecessary model complexity and risk overfitting, especially when domain knowledge suggests a well-defined temporal shape.

In contrast, an emerging class of models focuses on fitting parametric unimodal functions - such as the generalized logistic (e.g., Richards) \cite{tjorve2010unified,wang2012richards,aljarrah2020generalized,cortes2024weighted,stroup2024generalized}, skew-normal \cite{azzalini2005skew,zadkarami2010application,lee2013mixtures,shafiei2016alpha,said2017likelihood}
 or generalized gamma \cite{cordeiro2011exponentiated,li2011empirical,bourguignon2015new,thurai2018application,lahcene2021new} - to describe the evolution of a univariate quantity over time. These functions are inherently designed to capture bell-shaped or skewed temporal profiles with a single peak, making them particularly well-suited for phenomena such as seasonal biological activity, pollutant concentrations, or temperature responses, where a rise-and-fall pattern is expected.

The key advantage of such parametric approaches lies in their simplicity and parsimony. By reducing a time series to a small number of interpretable parameters (e.g., peak time, amplitude, spread, skewness), these models provide compact summaries of temporal behavior. This stands in contrast to decision trees or symbolic models, which may require dozens of nodes or terms to achieve similar predictive performance for simple unimodal dynamics.

Moreover, the novelty of these unimodal function-based models lies in their potential for domain-specific interpretability and model regularization. Parameters can often be directly mapped to physical or biological concepts (e.g., growth rates, onset timing), making them highly suitable for scientific applications where insight into temporal dynamics is as important as prediction. These models also offer robust performance in low-data regimes and naturally handle extrapolation, especially near the tails, where data-driven methods may struggle.

Another benefit is computational efficiency. Parametric curve fitting is typically orders of magnitude faster than training complex machine learning models and can be implemented with standard optimization techniques. This makes unimodal function models ideal for applications requiring real-time analysis, automated monitoring, or integration into embedded systems. While complex machine learning models excel in high-dimensional and relational settings, the use of unimodal parametric functions offers a lightweight, interpretable, and domain-aligned alternative for modeling univariate time series with a single prominent peak.

\section{Results and Discussion}

Building on the advantages of parametric unimodal functions, an additional compelling approach arises from information theory: using functions derived from the maximum entropy principle as candidate models for unimodal time series. The principle of maximum entropy selects the probability distribution that best represents the current state of knowledge while making the fewest assumptions beyond the given constraints - typically moments like the mean, variance, or known support. When applied to temporal modeling, this leads to least-biased functional forms that encode only what is known (e.g., the existence of a single peak, finite time support, skewness) and nothing more.

We aim to identify candidate functions on a bounded interval that maximize the entropy functional. Without loss of generality we can consider the bounded interval to be $[0, 1]$. Let $p(x)$ be a function that maximizes the entropy functional:
\begin{equation}
    H = -\int_{0}^{1} p(x) \log p(x) dx
\label{eq:1}
\end{equation}
under the following constraints:
\begin{equation}
\int_{0}^{1} p(x) dx = C_{1}\\
\label{eq:2}
\end{equation}
and
\begin{equation}
p(0) = 0 \mbox{ and } p(1) = 0
\label{eq:3}
\end{equation}

where $C_{1}$ is a finite constant (typically $C_{1} = 1$ for probability distributions). We can encode the constraints in \eqref{eq:3} by considering functions $f(x), g(x)$ such that $\lim_{x \rightarrow 0} |f(x)| = \infty$ and $\lim_{x \rightarrow 1} |g(x)| = \infty$ and enforcing the following properties:
\begin{equation}
\lim_{\epsilon \rightarrow 0}\int_{\epsilon}^{1} f(x)p(x) dx = C_{2} \mbox{ and } \lim_{\epsilon \rightarrow 0}\int_{0}^{1-\epsilon} g(x)p(x) dx = C_{3}
\label{eq:4}
\end{equation}
where $C_{2}$ and $C_{3}$ are also finite constants. The conditions in \eqref{eq:4}, together with suitable choices of $f(x)$ and $g(x)$, can enforce that $p(x)\to 0$ as $x\rightarrow 0$ or $1$, which are the desired constraints in \eqref{eq:3}.

Using Lagrange multipliers,  we can maximize the entropy functional $H$ under the constraints imposed in \eqref{eq:2} and \eqref{eq:4}. Different choices of functions $f(x)$ and $g(x)$ lead to different maximum entropy models. If $f(x) = \log{x}$ and $g(x) = \log{(1-x)}$ then the solution to the Euler-Lagrange equation derived from \eqref{eq:1} with constraints \eqref{eq:2} and \eqref{eq:4} is (up to a normalization constant): 
\begin{equation}
    p(x) \sim x^{a - 1} (1-x)^{b-1}
    \label{eq:beta}
\end{equation}
where $a > 1$ and $b > 1$ are parameters. Upon normalization this yields the $\beta(a, b)$ distribution, which is a well-known, classic probability distribution. For specific parameter choices (such as $a = 1 , b = 1$), the $\beta$ distribution is known to correspond to a maximum entropy distribution (e.g., the uniform distribution). However, we have shown that the beta distribution is generally a maximum entropy distribution if the constraints \eqref{eq:4} are imposed.

If $f(x) = \cfrac{1}{x}$ and $g(x) = \cfrac{1}{1-x}$ then:
\begin{equation}
    p(x) \sim \exp{\left(-\frac{a}{x}-\frac{b}{1-x}\right)}
\label{eq:MaxEnt}
\end{equation}
where $a, b > 0$ are parameters. We refer to \eqref{eq:MaxEnt} as the maximum entropy (MaxEnt) function (or distribution). Its mode (peak) is located at $x = \cfrac{\sqrt{a}}{\sqrt{a}+\sqrt{b}}$. While other choices of maximum entropy models are possible, we focus in this paper on applications of \eqref{eq:beta} and \eqref{eq:MaxEnt}.

\setlength\tabcolsep{1em}
\begin{table}[t]
\Huge
\centering
    \resizebox{\textwidth}{!}{ 
\begin{tabular}{l|c|c|c|c|c}
Methods    & Richards        & Skewnormal      & GenGamma        & MaxEnt           & Beta            \\
\hline
Richards   & 0.0018(98) & 0.04(7)    & 0.05(8)   & 0.04(1)    & 0.08(5)   \\
\hline
Skewnormal & 0.12(8)    & 0.07(4)   & 0.3(1)      & 0.28(6)     & 0.47(8)    \\
\hline
GenGamma   & 0.31(9)    & 0.33(3)    & 0.4(1)     & 0.4(1)      & 0.3(2)     \\
\hline
MaxEnt     & \textbf{0.03(2)}    & \textbf{0.03(1)}   & 0.06(9)   & 0.0002(30)& \textbf{0.05(10)}   \\
\hline
Beta       & \textbf{0.03(2)}   & 0.03(2)   & \textbf{0.03(2)}    & \textbf{0.011(5)}   & 0.001(3) \\
\end{tabular}
}
\vspace{2mm}
\caption{\small Root Mean Squared (RMS) loss and standard deviations from cross-comparison of different unimodal models used to fit synthetic time series generated by each method. Diagonal values represent each model fitting its own data, while off-diagonal values highlight cross-model fitting performance.}
\vspace{-7mm}
\label{tab:1}
\end{table}

\subsection{Method comparison}

Table \ref{tab:1} presents a cross-comparison of five different methods used to model unimodal time series: Richards, Skewnormal, Generalized Gamma (GenGamma), Maximum Entropy (MaxEnt), and Beta functions. Each row corresponds to the method used for fitting, while each column corresponds to the method that generated the synthetic time series data. The values represent the root mean squared loss in fit accuracy, with standard deviations shown in parentheses. Lower values indicate better performance.

Richards, Skewnormal, and Generalized Gamma functions form the core set of reference models in our comparative analysis because they offer flexible representations of asymmetry - a crucial feature for accurately modeling unimodal time series. Each of these functions can express both left-skewed and right-skewed shapes, allowing them to accommodate a wide range of peak timings and growth-collapse dynamics. The Richards function, as a generalized logistic model, adjusts its curvature through a shape parameter that governs asymmetry. The Skewnormal distribution introduces skew through a shape parameter modifying the Gaussian form, while the Generalized Gamma offers even greater flexibility by encompassing a family of distributions that include both symmetric and highly skewed profiles. These three models serve as the primary benchmarks against which we evaluate the performance of the two-parameter maximum entropy functions.

In contrast, other distributions commonly used in practice - such as the Lognormal and Gompertz - exhibit important limitations. The Lognormal is inherently right-skewed, making it unsuitable for processes with early peaks or left-skewed dynamics. The Gompertz function is in fact a special case of the Richards curve, and as such, it is implicitly included in our comparison through the Richards model itself.

As expected, the diagonal entries - where the model is applied to data generated by the same functional form - in general show the best performance across the board. This is intuitive: each method is best suited to reproducing its own characteristic shape and temporal dynamics. For instance, Richards performs best on Richards-generated data (0.0018), and similarly for the Skewnormal (0.07), MaxEnt (0.0002), and Beta (0.001) models. GenGamma is an exception, showing similarly poor performance when fitting both its own data and data from other models. This suggests that, contrary to its theoretical flexibility, the model may be more rigid in practice, or its parameter estimation may be especially prone to becoming trapped in local minima during optimization. As a result, GenGamma fails to capture its own characteristic dynamics as effectively as the other methods.

What is more revealing are the off-diagonal values, particularly the second-best scores, which highlight the versatility of models in capturing other functional forms. These are emphasized in bold within the table. A key finding from this analysis is the consistently strong cross-fit performance of the MaxEnt model. When applied to time series generated by other methods, MaxEnt achieves the lowest off-diagonal loss in three out of four cases: Richards (0.03), Skewnormal (0.03), and Beta (0.05). This suggests that the MaxEnt-derived models, which are based on the principle of maximizing entropy under shape constraints, provide a highly flexible and unbiased framework for modeling diverse unimodal time series shapes.

As we have shown, the Beta model is also a maximum entropy model and performs competitively in cross-fits, notably yielding the second-best loss on GenGamma-generated data (0.03) and a strong performance on MaxEnt-generated data (0.011). This reflects the Beta function's capacity to adapt to various peak asymmetries and supports its use in bounded time interval modeling.

In contrast, the GenGamma and Skewnormal models show poorer generalization to other types, with higher cross-fit errors - especially when fitting MaxEnt or Richards-generated data. In addition, the Richards, GenGamma and Skewnormal methods enjoy three parameters (or degrees of freedom) and should have a better capacity at fitting diverse patterns. However, the results in Table \ref{tab:1} illustrate that maximum entropy-based models offer the best overall versatility, capturing a wide range of unimodal dynamics with minimal loss. Their data-agnostic nature, guided only by moment constraints, makes them ideal candidates for general-purpose modeling where prior knowledge of the underlying process is limited.

\subsection{Universe 25}

Universe 25 is a behavioral experiment from the 1970s by American ethologist John B. Calhoun, designed to study the effects of overpopulation on social behavior in rodents \cite{calhoun1973death}. It involved a population of mice placed in a "utopia" — an enclosed space with unlimited food, water, nesting material, and no predators. The only limiting factor was space. The experiment aimed to simulate conditions of density-dependent stress.

Initially, the population grew rapidly, but over time, social breakdown emerged. Dominant males became aggressive, maternal behavior deteriorated, and many mice withdrew from social interaction entirely. Eventually, the population collapsed entirely, despite the continued presence of resources. This phenomenon, which Calhoun called the ``behavioral sink'', suggested that social and psychological stress from overcrowding could override even ideal environmental conditions.

\begin{figure}[t]
\centering
\includegraphics[width=\textwidth]{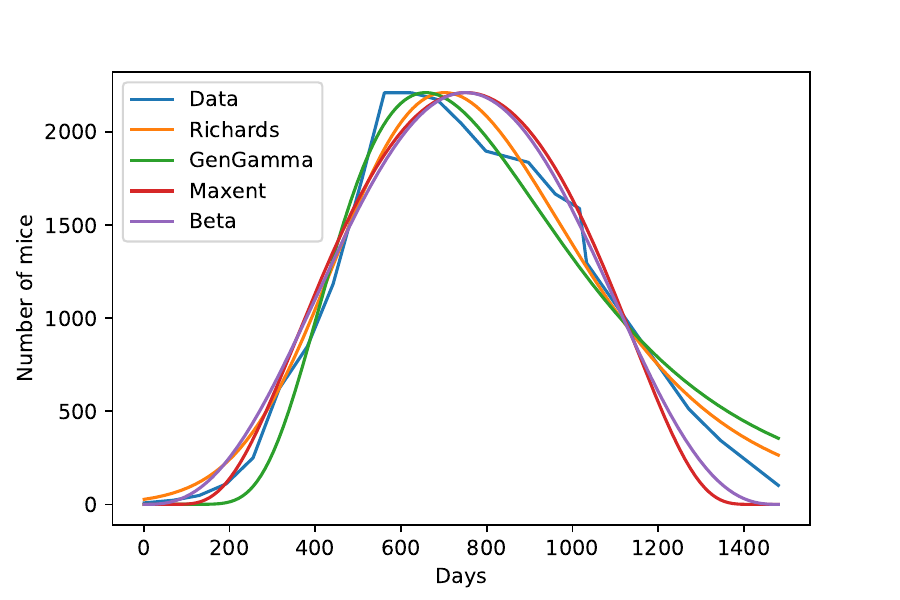}
\caption{\small Mouse population trajectory from the Universe 25 experiment, fitted with four unimodal models: Richards, GenGamma, MaxEnt, and Beta.}
\label{fig:1}
\vspace{-5mm}
\end{figure}

Figure \ref{fig:1} shows the mouse population curve over time from the Universe 25 experiment, plotted alongside fitted curves using five different unimodal modeling approaches: Richards, Generalized Gamma (GenGamma), Maximum Entropy (MaxEnt), and Beta functions. The population data exhibits a characteristic unimodal shape - a rapid growth phase, a population peak, followed by a decline and collapse. The Richards function is a sigmoid-derived model that captures the general shape well, particularly the early growth and peak region. However, it slightly underperforms in capturing the asymmetry of the decline, showing a more gradual descent than observed in the data. The Generalized Gamma distribution is flexible, but in this case, it overestimates the early growth and smooths over some of the sharper transitions in the decline. It offers a reasonable fit but appears to lag in accurately modeling the abrupt changes post-peak.

The MaxEnt-derived curve, based on maximizing entropy under moment constraints, shows a strong fit across the entire curve, especially in the peak and decline regions. It follows the asymmetric descent more closely than most others, making it a strong candidate for capturing the essential features of the population trajectory. The Beta function also fits well, particularly around the growth and peak regions.

The fitted models each provide unique insights into the population dynamics of Universe 25. While all manage to reproduce the general unimodal shape, MaxEnt and Beta functions stand out for their flexibility and fidelity to the real data, particularly in modeling the asymmetric shape of the rise and collapse. The MaxEnt model’s performance highlights the power of information-theoretic approaches in modeling complex natural phenomena with minimal assumptions, capturing both symmetry and skewness effectively. Overall, this figure supports the conclusion that modeling with flexible, peak-constrained functions like MaxEnt or Beta distributions can offer more accurate, data-consistent reconstructions of complex behavioral phenomena like those observed in Universe 25.

\subsection{Saint Matthew Island}

\begin{figure}[t]
\centering
\includegraphics[width=\textwidth]{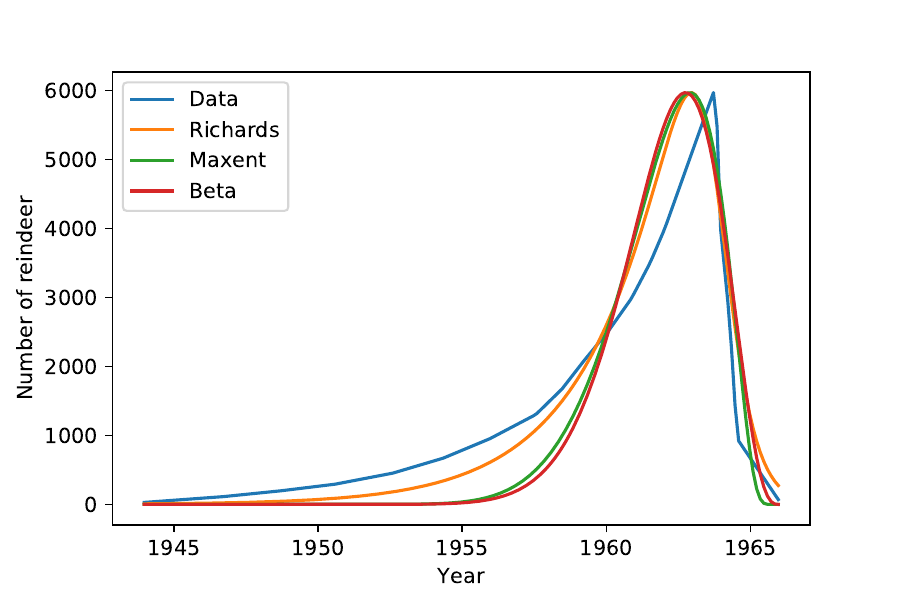}
\caption{\small Reindeer population on Saint Matthew Island (1944-1966) with fitted curves from three unimodal models: Richards, MaxEnt, and Beta. The data show rapid population growth followed by a sharp collapse.}
\label{fig:2}
\vspace{-5mm}
\end{figure}

Saint Matthew Island, a remote landmass in the Bering Sea, is home to one of the most dramatic cases of wildlife population collapse ever recorded. In 1944, 29 reindeer were introduced to the island by the U.S. Coast Guard as a food source for a wartime weather station crew \cite{klein1968introduction}. The station was soon abandoned, leaving the reindeer without predators or human interference. With abundant vegetation and no natural predators, the reindeer population grew exponentially. By 1963, their numbers had ballooned to approximately 6,000. However, the island’s environment could not sustain such a large population. Overgrazing led to severe habitat degradation, particularly of the slow-growing lichens that formed their primary winter food source. The population crashed catastrophically during the harsh winter of 1963–64. By 1966, only 42 reindeer remained—41 females and a single male—none of which were reproductively viable. Eventually, the population went extinct. Figure \ref{fig:2} illustrates the reindeer population dynamics on Saint Matthew Island from their introduction in the 1940s until their catastrophic collapse in the mid-1960s. The empirical data shows a slow initial growth followed by a rapid exponential increase, peaking around 1963, and then a sudden, steep collapse. 

As a generalized logistic function, the Richards curve captures the early and mid-phase population growth fairly well. However, it underestimates the sharpness of the collapse. Its symmetry and gradual decline make it less suited to modeling abrupt ecological breakdowns. The MaxEnt-derived curve shows good adaptability to the data, particularly in reflecting the asymmetrical shape of the population curve—rapid growth followed by an abrupt drop. This reflects the model’s strength in capturing real-world ecological constraints without assuming specific underlying mechanisms, making it well-suited for modeling systems with sudden transitions or tipping points. The Beta function also performs well, closely matching the data through the peak and collapse phase. Its bounded nature makes it inherently suitable for modeling finite-time processes, such as a population rising and falling within a constrained time frame.

The extreme dynamics of the Saint Matthew Island reindeer population are a classic example of overshoot and collapse due to resource overexploitation. The ability of MaxEnt and Beta models to closely fit this trajectory underscores their usefulness in ecological modeling, especially when dealing with systems prone to nonlinear behavior and irreversible transitions. These models, grounded in shape and peak constraints, provide a powerful alternative to traditional growth models, offering better sensitivity to abrupt ecological changes such as extinction or collapse events.

\section{Conclusion}

This study presents a comparative analysis of various unimodal models - Richards, Skewnormal, Generalized Gamma, Beta, and a novel class of Maximum Entropy (MaxEnt) functions - for modeling time series that exhibit a single peak followed by a decline. Through both synthetic data experiments and real-world case studies, including the Universe 25 mouse population collapse and the Saint Matthew Island reindeer die-off, we evaluated each model's ability to fit and generalize complex unimodal dynamics.

A key result from the cross-comparison table of RMS losses is that, while each model performs best when fitting data generated from its own functional form, the maximum entropy models (Beta and MaxEnt) consistently achieve the lowest off-diagonal loss values. This indicates superior flexibility and generalization across different types of unimodal behavior. Importantly, the maximum entropy models accomplish this with only two degrees of freedom, in contrast to the three-parameter structure of the other models evaluated. This makes them not only more parsimonious but also less prone to overfitting, a critical advantage when data are sparse or noisy.

The empirical fits further support this conclusion. In both the Universe 25 and Saint Matthew Island datasets, the maximum entropy models offered closer tracking of asymmetric growth and collapse patterns, outperforming traditional models that tend to impose more symmetric or smoothed shapes. The Beta function is also competitive in accuracy and shares a similar peak-constrained form.

Taken together, these results demonstrate that maximum entropy-based models represent a novel, principled, and efficient framework for modeling unimodal time series. Their ability to provide strong fits using only shape-based constraints - without reliance on parametric assumptions or domain-specific tuning - makes them especially promising for applications in ecology, epidemiology, demography, and beyond. They serve not only as accurate approximators of known dynamics but also as least-biased candidates when functional forms are unknown or poorly understood.

\section*{Data Availability Statement}
The code to reproduce the results is available at:\\
\url{https://doi.org/10.5281/zenodo.16088580}

\section*{Conflict of Interest Statement}

The authors declare that the research was conducted in the absence of any commercial or financial relationships that could be construed as a potential conflict of interest.

\begin{credits}
\subsubsection{\ackname}

This publication is supported by the European Union's Horizon Europe research and innovation programme under the Marie Sk\l{}odowska-Curie Postdoctoral Fellowship Programme, SMASH co-funded under the grant agreement No. 101081355. The operation (SMASH project) is co-funded by the Republic of Slovenia and the European Union from the European Regional Development Fund.

\end{credits}
%
%
%
\bibliographystyle{splncs04}
\bibliography{mybib}

\end{document}